\documentclass[preprint,12pt]{elsarticle}

\usepackage{a4wide}
\usepackage{amsmath,amssymb}
\usepackage{wasysym}
\usepackage{graphics}
\usepackage{latexsym}
\usepackage[usenames]{color}

  \usepackage{macros_art}

\usepackage{MnSymbol}

\usepackage{tikz}
\usetikzlibrary{shapes}

\usepackage{soul}

\newtheorem{thm}{Theorem}
\newtheorem{claim}{Claim}
\newtheorem{remark}{Remark}
\newproof{pf}{Proof}

\def\cP{\mathcal{P}}
\def\cG{\mathcal{G}}
\def\cT{\mathcal{T}}

\def\ra{\rightarrow}
\def\Ra{{\Rightarrow}}
\def\Lra{{\Leftrightarrow}}
\def\sdpart{\rightharpoonup}

\newcommand{\ceoal}[1][A]{{\llangle}{\mathit #1}{\rrangle}}
\newcommand{\ceggg}[1][A]{\ceoal[#1]\!\Box}
\newcommand{\cenxt}[1][A]{\ceoal[#1]\!\!\ocircle\!}

\begin{document}

\begin{frontmatter}

\title{Model-checking $ATL$ under Imperfect Information 
       and Perfect Recall Semantics is Undecidable}

\author[a]{C\u at\u alin Dima}
\ead{dima@univ-paris12.fr}

\author[b]{Ferucio Lauren\c tiu \c Tiplea}
\ead{fltiplea@info.uaic.ro}

\address[a]{LACL, Universit\'e Paris Est-Cr\'eteil, 
                 61 av. du G-ral de Gaulle, 
                 94010 Cr\'eteil, France}
                 
\address[b]{Department of Computer Science, 
                 ``Al.I.Cuza'' University of Ia\c si,
                 Ia\c si 700506, Romania}

\begin{abstract}
We propose a formal proof of the undecidability
of the model checking problem for alternating-time temporal
logic under imperfect information and perfect recall semantics.
This problem was announced to be undecidable according
to a personal communication on multi-player games with
imperfect information, but no formal proof was ever published.
Our proof is based on a direct reduction from the non-halting
problem for Turing machines. 
\end{abstract}


\end{frontmatter}

\section{Introduction}

The Alternating-time Temporal Logic ($ATL$) have been introduced
in \cite{AlHK2002} as a logic to reason about strategic abilities 
of agents in multi-agent systems. $ATL$ extends $CTL$ by replacing
the path quantifiers $\forall$ and $\exists$ by \emph{cooperation
modalities} $\ceoal[A]$, where $A$ is a team of agents. A formula
$\ceoal[A]\varphi$ expresses that the team $A$ has a collective
strategy to enforce $\varphi$. 

The semantics of $ATL$ is defined over 
{\em concurrent game structures} ($CGS$) \cite{AlHK2002} 
which are transition systems whose states are labeled
by atomic propositions and for which a set of agents is
specified. Each agent may have {\em incomplete/imperfect
information} about the state of the system in the sense 
that the agent 
may not be able to difference between some states.
When the agent is able
to observe the entire state labeling, we say that he
has {\em complete/perfect information}. 
A transition from a state to another one is performed by 
an action tuple consisting of an action for each agent in 
the system. The action an agent is allowed to 
perform at a state is chosen from a given set of actioned
allowed to be performed by the agent at that state and may
depend on the current state (this is called {\em imperfect recall\/})
or on the whole history of events that have happened
(this is called {\em perfect recall\/}). 
Combining imperfect or perfect information with imperfect
or perfect recall we obtain four types of concurrent game
structures and, consequently, four types of semantics for
$ATL$. 

A series of papers have been addressed the model-checking 
problem for $ATL$ \cite{AlHK2002,Scho2004,BuDJ2010}.
Based on unpublished work of Yannakakis \cite{Yann1997}, 
the model checking problem for $ATL$ with imperfect information
and perfect recall semantics was announced to be undecidable
in \cite{AlHK2002}. Since then, many authors have mentioned
this result but, unfortunately, no formal proof was ever 
published (see also \cite{BuDJ2010}). 

In this paper we propose a formal proof of this problem. 
Our proof is based on a direct simulation of Turing machines
by concurrent game structures under imperfect information
and perfect recall, which allows for a reduction of the
non-halting problem for Turing machines to the model checking
problem for $ATL$ under imperfect information and perfect 
recall semantics. 
Moreover, the strategies used by agents to simulate the
Turing machine are primitive recursive. This shows that 
the undecidability of model checking $ATL$ under imperfect 
information and perfect recall semantics is mainly due to
the imperfect information agents have about the system
states.  

While our proof is given for the \emph{de dicto} strategies
from \cite{AlHK2002}, the same construction 
works also for the \emph{de re} strategies from 
\cite{Scho2004,JaAg2007}.

\section{Alternating-time Temporal Logic}\label{S2}

We recall in this section the syntax and semantics of the
alternating-time temporal logic. We will mainly follow 
the approach in \cite{BuDJ2010} and fix first a few notations.
$\Nset$ stands for the set of positive integers (natural
numbers) and $\PPP$ denotes the powerset operator.
Given a set $V$, $V^+$ denotes the free semi-group and $V^*$ 
denotes the free monoid generated by $V$ under concatenation. 
$\lambda$ stands for the empty word (the unity of $V^*$). 
The notation $f:X\sdpart Y$ means that $f$ is a partially 
defined function from $X$ to $Y$.

\paragraph{$ATL$ syntax}

The syntax of $ATL$ is given by the grammar
  $$\varphi::= p \mid \neg\varphi \mid \varphi\wedge\varphi 
                 \mid \cenxt[A]\varphi 
                 \mid \ceggg[A]\varphi 
                 \mid \ceoal[A]\varphi U\varphi$$
where $p$ ranges over a finite non-empty set of {\em atomic propositions}
$\Pi$, $A$ is a non-empty subset of a finite set $Ag$ of {\em agents},
and $\ocircle$, $\Box$, and $U$ are the standard temporal operators
{\em next}, {\em globally}, and {\em until}, respectively. 

Note that, in order to define combinations of temporal operators 
inside the coalition operators, the \emph{weak-until} operator
should be given as a primitive operator \cite{LaMO2008}, since it 
cannot be derived from the above operators.
However our result holds also for this restricted syntax.

\paragraph{$ATL$ semantics}

$ATL$ is interpreted over {\em concurrent game structures} ($CGS$)
\cite{AlHK2002}. 
Such a structure consists of a set of states labeled by atomic 
propositions and a set of agents. Each agent may perform some
actions and at least one action is available to the agent at each state. 
His decision in choosing which action should be performed at some state 
may be based on his capability of observing all or some atomic 
propositions at the current state, usually called {\em perfect} 
or {\em imperfect information}, and on his full or partial history, 
usually called {\em perfect} or {\em imperfect recall}. 

In what follows we focus on $CGS$ under imperfect information
and perfect recall and adopt the formal approach in \cite{BuDJ2010}.
A {\em $CGS$ under imperfect information} is a tuple 
$\cG=(Ag,S,\Pi,\pi,(\sim_i|i\in Ag),Act,d,\ra)$, where:
\begin{itemize}
\item $Ag=\{1,\ldots,k\}$ is a finite non-empty set of {\em agents};
\item $S$ is a finite non-empty set of {\em states}; 
\item $\Pi$ is a finite non-empty set of {\em atomic propositions}; 
\item $\pi:S\ra\cP(\Pi)$ is the {\em state-labeling function};
\item $\sim_i$ is an equivalence relation on $S$, for any
      agent $i$;
\item $Act$ is a finite non-empty set of {\em actions}; 
\item $d:Ag\times S\ra \cP(Act)-\{\emptyset\}$ gives the set of 
      actions available to agents at each state, satisfying
      $d(i,s)=d(i,s')$ for any agent $i$ and states $s$ and
      $s'$ with $s\,\sim_i\,s'$;
\item $\ra:S\times Act^{k}\sdpart S$ is the (partially defined)
      {\em transition function} satisfying, for any $s\in S$ and 
      $(a_1,\ldots,a_k)\in Act^k$, the following property:
        $$\ra(s,(a_1,\ldots,a_{k}))\ \mbox{is defined iff }
         a_i\in d(i,s)\ \mbox{for any agent $i$}.$$
      We will write $s\sdup{(a_1,\ldots,a_k)}s'$, whenever 
      $\ra(s,(a_1,\ldots,a_{k}))=s'$.
\end{itemize}

If $s$ and $s'$ are $\sim_i$-equivalent (i.e., $s\,\sim_i\,s'$)
then we say that $s$ and $s'$ are {\em indistinguishable} from
the agent $i$'s point of view (due to his partial view on the states).
Each $\sim_i$ is component-wise extended to sequences of
states. 
Thus, for $\alpha,\alpha'\in S^+$ we write $\alpha\sim_i\alpha'$
and say that $\alpha$ and $\alpha'$ are {\em $\sim_i$-equivalent}
if $\alpha=s_0\cdots s_n$ and $\alpha'=s_0'\cdots s_n'$ for some
$n\in\Nset$, and $s_j\sim_i s_j'$ for all $0\leq j\leq n$.

A {\em perfect recall strategy} for
an agent $i$ in a $CGS$ $\cG$ 
is a function $\sigma:S^+\ra Act$ which is {\em compatible} with 
$d$ and $\sim_i$, i.e., 
\begin{itemize}
\item $\sigma(\alpha s)\in d(i,s)$, for any 
  $\alpha\in S^*$ and $s\in S$;
\item $\sigma(\alpha)=\sigma(\alpha')$, for
  any $\alpha,\alpha'\in S^+$ with $\alpha\,\sim_i\,\alpha'$.
\end{itemize}

A {\em perfect recall strategy for a team $A$ of agents} is a
family $\sigma_A=(\sigma_i|i\in A)$ of perfect recall strategies 
for the agents in $A$. 
If $\sigma_A$ is a perfect recall strategy for the agents in 
$A$, $\alpha s\in S^*S$, and $a=(a_1,\ldots,a_k)\in Act^k$, 
then we write $a\in\overline{\sigma}_A(\alpha s)$ if the 
following properties hold:
\begin{itemize}
\item $a_i\in d(i,s)$, for any $i\in Ag-A$;
\item $a_i\in\sigma_i(\alpha s)$, for any $i\in A$.
\end{itemize}

Given a state $s$ of $\cG$ and $\sigma_A$ as above, define
$out_\cG(s,\sigma_A)$ as being the set of all infinite
sequences of states $\lambda=s_0s_1s_2\cdots$ such that 
$s_0=s$ and, for any $j\geq 0$, there exists 
$a\in\overline{\sigma}_A(s_0\cdots s_j)$ with 
$s_j\sdup{a}s_{j+1}$.
For $\lambda=s_0s_1s_2\cdots$ an infinite sequence of states 
and $j\geq 0$, $\lambda[j]$ denotes the $j$-th state in the 
sequence, $\lambda[j]=s_j$

The {\em imperfect information perfect recall semantics} for
$ATL$, denoted $\models_{iR}$, is defined as follows
($\cG$ is a $CGS$ under imperfect information and $s$ is
a state of $\cG$):
\begin{itemize}
\item $(\cG,s)\models_{iR} p$ if $p\in\pi(s)$;
\item $(\cG,s)\models_{iR}\neg\varphi$ if $(\cG,s)\not\models_{iR}\varphi$;
\item $(\cG,s)\models_{iR}\varphi\wedge\psi$ if 
      $(\cG,s)\models_{iR}\varphi$ and $(\cG,s)\models_{iR}\psi$;
\item $(\cG,s)\models_{iR}\cenxt[A]\varphi$ if there exists a 
      perfect recall strategy $\sigma_A$ such that 
      $(\cG,\lambda[1])\models_{iR}\varphi$, for any 
      $\lambda\in out_\cG(s,\sigma_A)$;
\item $(\cG,s)\models_{iR}\ceggg[A]\varphi$ if there exists a 
      perfect recall strategy $\sigma_A$ such that 
      $(\cG,\lambda[j])\models_{iR}\varphi$, for any 
      $\lambda\in out_\cG(s,\sigma_A)$ and any $j\geq 0$;
\item $(\cG,s)\models_{iR}\ceoal[A]\varphi U\psi$ if there exists a 
      perfect recall strategy $\sigma_A$ such that for any
      $\lambda\in out_\cG(s,\sigma_A)$ there exists $j\geq 0$
      with $(\cG,\lambda[j])\models_{iR}\psi$ and
      $(\cG,\lambda[k])\models_{iR}\varphi$ for all
      $0\leq k<j$.
\end{itemize}

The {\em model checking problem} for $ATL$ formulas under 
imperfect information and perfect recall semantics is to
decide, given an $ATL$ formula $\varphi$, a concurrent game 
structure $\cG$ under imperfect information, and a state 
$s$ of $\cG$, whether $(\cG,s)\models_{iR}\varphi$.

\paragraph{Computation trees}

The proof of our main result in the next section will be based 
on {\em computation trees} associated to $CGS$s. These are
special cases of labeled trees, which are structures 
$\cT=(V,E,v_0,l_1,l_2)$, where 
\begin{itemize}
\item $(V,E,v_0)$ is a tree whose set of nodes is $V$,
  whose set of edges is $E$, and whose root is $v_0$;
\item $l_1$ is the node-labeling function;
\item $l_2$ is the edge-labeling function.
\end{itemize}

{\em Paths} in a labeled tree $\cT=(V,E,v_0,l_1,l_2)$ are
defined inductively as usual as sequences of nodes:
\begin{itemize}
\item $v_0$ is a path in $\cT$;
\item if $v_0\cdots v_n$ is a path in $\cT$ and $(v_n,v)\in E$,
  then $v_0\cdots v_nv$ is a path in $\cT$.
\end{itemize}
If $v$ is a node of $\cT$, then $path_\cT(v_0,v)$ stands
for the unique path from the root $v_0$ to $v$ in $\cT$.
The number of nodes on a path $\tau$ is the length of $\tau$,
denoted $|\tau|$. 
The labeling function $l_1$ is homomorphically extended to paths,
that is, $l_1(\tau_1\tau_2)=l_1(\tau_1)l_1(\tau_2)$.

{\em Levels} in a labeled tree $\cT=(V,E,v_0,l_1,l_2)$ are
sets of nodes of $\cT$ defined inductively as follows:
\begin{itemize}
\item $level_\cT(0)=\{v_0\}$;
\item $level_\cT(n+1)=\{v\in V|
       (\exists v'\in level_\cT(n))((v',v)\in E)\}$,
      for any $n\geq 0$.
\end{itemize}
$level_\cT(n)$ is referred to as the {\em level $n$ in $\cT$}.

Given a CGS $\cG$, a state $s$ of $\cG$, a coalition $A$
of agents, and a perfect recall strategy $\sigma_A$ for
agents in $A$, define inductively the 
{\em $s$-rooted computation trees of $\cG$ under $\sigma_A$} 
as follows:
\begin{itemize}
\item any tree with exactly one node (its root) labeled by $s$
  is an $s$-rooted computation tree of $\cG$ under $\sigma_A$;
\item if $\cT=(V,E,v_0,l_1,l_2)$ is an $s$-rooted computation 
  tree of $\cG$ under $\sigma_A$, $v$ is a node of $\cT$, and 
  $l_1(v)\sdup{a} s'$ for some action-tuple 
  $a\in\overline{\sigma}_A(l_1(path_\cT(v_0,v)))$ and state $s'$ 
  such that no edge from $v$ is labeled by $a$, then the tree $\cT'$ 
  obtained as follows is an $s$-rooted computation tree of $\cG$:
  \begin{itemize}
  \item $\cT'$ is obtained from $\cT$ by adding a new node $v'$
    labeled by $s'$ and an edge $(v,v')$ labeled by $a$.
  \end{itemize}
\end{itemize}
If $\cT'$ is obtained from $\cT$ as above, we will also write
$\cT\Ra_{\cG,\sigma_A}\cT'$ or 
$\cT\stackrel{a}{\Ra}_{\cG,\sigma_A}\cT'$ if we want to specify the
action tuple $a$ as well.

\begin{remark}\label{R1}
It is easy to see that, for any atomic proposition $p$, the 
following property holds true:
\begin{itemize}
\item $(\cG,s)\models_{iR}\ceggg[A]p$ if and only if there 
      exists a perfect recall strategy $\sigma_A$ such that 
      $p\in\pi(l_1(v))$, for any $s$-rooted computation tree 
      $\cT$ of $\cG$ under $\sigma_A$, and any node $v$ of
      $\cT$.
\end{itemize}
\end{remark}

\section{Undecidability of Model Checking $ATL_{iR}$}

We will prove in this section that the model checking
problem for $ATL_{iR}$ is undecidable. 
The proof technique is by reduction from the non-halting 
problem for deterministic Turing machines. 
Given a deterministic Turing machine $M$, 
we construct a concurrent game structure under imperfect information 
$\cG$ with three agents $Ag=\{1,2,3\}$, a state 
$s_{init}$ of $\cG$, and an $ATL$ formula 
$\ceggg[\{1,2\}]ok$, where $ok$ is an atomic 
proposition, such that $M$ does not halt on the empty 
word if and only if 
$(\cG,s_{init})\models_{iR}\ceggg[\{1,2\}]ok$.

The deterministic Turing machines we consider are tuples
$M=(Q,\Sigma,q_0,B,\delta)$, where $Q$ is a finite set 
of states, $\Sigma$ is a finite tape alphabet, $q_0$ is 
the initial state, $B\in\Sigma$ is the blank symbol, and 
$\delta:Q\times\Sigma\sdpart Q\times\Sigma\times\{L,R\}$ 
is a partially defined transition function, where ``$L$'' 
specifies a ``left move'' and ``$R$'' specifies a ``right move''.
A configuration of $M$ is a word $a_1\cdots a_{i-1}qa_i\cdots a_n$,
where all $a$'s are from $\Sigma$ and $q$ is a state.
Such a configuration specifies that $M$ is in state $q$, its
read/write head points to the $i$th cell of the tape, and the 
$j$th cell holds $a_j$ if $j\leq n$, and $B$, otherwise. 
The initial configuration is $q_0B$. 
The transition relation on configurations, denoted $\Ra_M$,
is defined as usual. For instance,
$a_1\cdots a_{i-1}qa_i\cdots a_n\Ra_M a_1\cdots q'a_{i-1}a_i'\cdots a_n$
if $i>1$ and $\delta(q,a_i)=(q',a_i',L)$.

The Turing machine $M$ {\em halts} on the empty word if,
starting with the initial configuration, the machine 
reaches a configuration $a_1\cdots a_{i-1}qa_i\cdots a_n$
for which $\delta(q,a_i)$ is undefined or $i=1$ and
$\delta(q,a_i)=(q',a_i',L)$ for some $q'$ and $a_i'$.

\paragraph{Intuition first}

\begin{figure}[hbt]
\centering
\begin{tikzpicture}
[level distance=1.5cm,
 level 1/.style={sibling distance=1.5cm},
 every node/.style={circle,draw,minimum size=.5cm},
 every text node part/.style={font=\small},
 every lower node part/.style={font=\small}]
 \node[circle,draw] {$s_{init}$}
   child {
     node {$s_{lb}'$} 
     edge from parent[dashed]
          }
   child {
     node {$s_{a_1}$} 
     edge from parent[dashed]
          } 
   child {
     node {$s_{tr}'$} 
     edge from parent[dashed]
          } 
   child {
     node {$s_{q,a_2}$}
     edge from parent[dashed]
          } 
   child {
     node {$s_{tr}'$} 
     edge from parent[dashed]
          } 
   child {
     node {$s_{a_3}$} 
     edge from parent[dashed]
          };
\end{tikzpicture}
\caption{Level corresponding to $a_1qa_2a_3$}
\label{F1}                                
\end{figure}
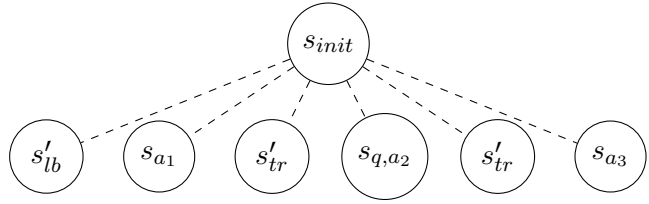

The main idea of the construction is to encode the configurations 
of the Turing machine horizontally in the levels of the computation 
tree.
A configuration $a_1\cdots a_{i-1}qa_i\cdots a_k$ of $M$
will be simulated in $\AAA$ by some level in some 
computation tree like in Figure \ref{F1} (where $i=2$ and $k=3$).
The nodes of this tree are represented by circles. The label
of a node is carried inside the circle representing the node. 
The node labeled $s_{lb}'$ specifies the left border of $M$'s 
tape, the node labeled $s_{tr}'$ is a cell separator also used 
to transfer information between paths of computation trees, 
the nodes labeled $s_{a_1}$ and $s_{a_3}$ specify the content 
of the first and third cell, respectively, 
and the node labeled $s_{q,a_2}$ specifies both the content of 
the second cell and the fact that $M$ is in state $q$ and its 
read/write head points to the second cell. 

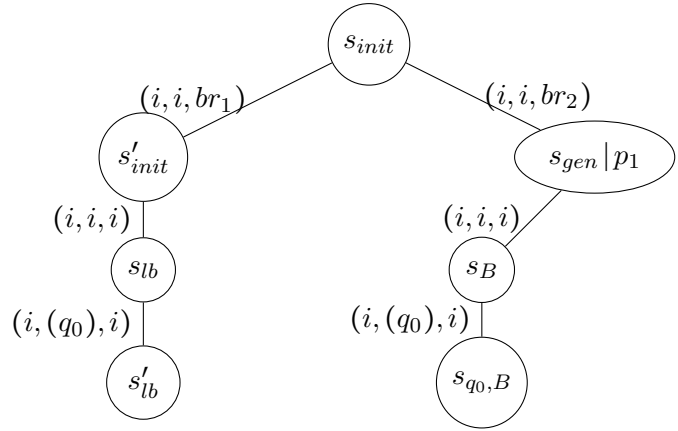
\begin{figure}[hbt]
\centering
\begin{tikzpicture}
[level distance=1.5cm,
 level 1/.style={sibling distance=6cm},
 level 2/.style={sibling distance=3cm},
 every text node part/.style={font=\small},
 every lower node part/.style={font=\small}]
 \node[circle,draw] {$s_{init}$}
   child {
     node[circle,draw] {$s_{init}'$}
     child {
       node[circle,draw] {$s_{lb}$}
       child {
         node[circle,draw] {$s_{lb}'$}
         edge from parent node[midway,left] {$(i,(q_0),i)$}
              }
       edge from parent node[midway,left] {$(i,i,i)$}
            } 
     edge from parent node[midway,left] {$(i,i,br_1)$}
          }
   child {
     node[ellipse,draw] {$s_{gen}\,|\,p_1$} 
     child {
       node[circle,draw] {$s_B$}
       child {
         node[circle,draw] {$s_{q_0,B}$}
         edge from parent node[midway,left] {$(i,(q_0),i)$}
              }
       edge from parent node[midway,left] {$(i,i,i)$}
            } 
     child[missing] {node {a}}     
     edge from parent node[midway,right] {$(i,i,br_2)$}
          };
\end{tikzpicture}
\caption{Generating the initial configuration $q_0B$ of $M$}
\label{F2}                                
\end{figure}

The generation of the initial configuration $q_0B$ of $M$ is 
simulated by the computation tree in Figure \ref{F2}. All 
states in this tree ale labeled by $ok$; the node labeled $s_{gen}$ 
has one more label, namely $p_1$ (this label is graphically 
represented because it will be particularly important in 
defining the agents strategies).
As we will see later, the two maximal paths in this tree 
are $\sim_2$-equivalent. 
This allows, together with the strategy we will use, for the 
synchronization in the last computation step of these paths. 

The levels encoding configurations of the Turing machine will 
be encoded on the \emph{even} positions in a computation tree,
the odd levels being used for correctly representing transitions 
of the Turing machine. Some nodes in the levels of even index 
will then encode tape cells, while some other nodes will be 
used for transferring information between adjacent cells.
Some examples presenting this idea are given in the following, 
before the formal construction and proof.

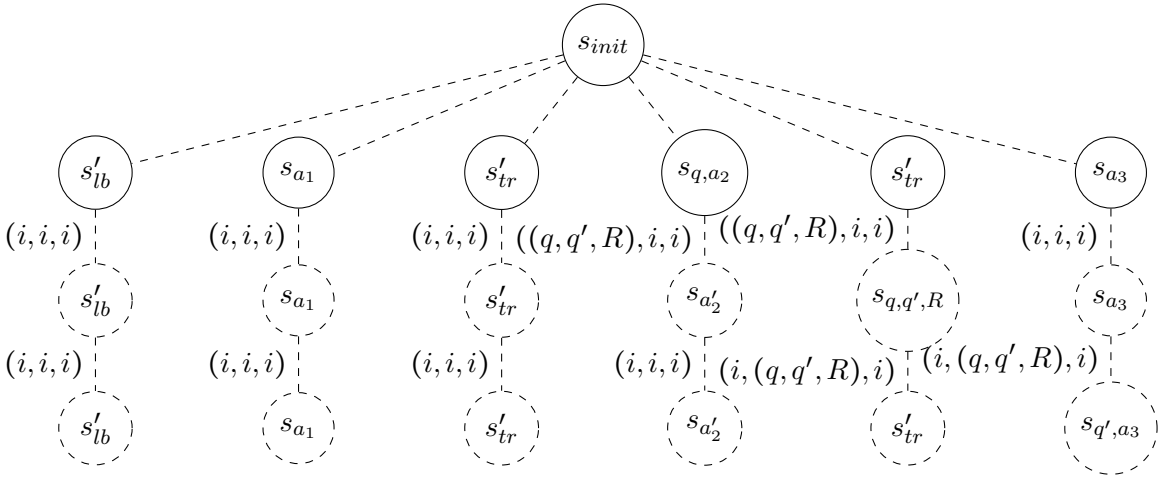
\begin{figure}[hbt]
\centering
\begin{tikzpicture}
[level distance=1.7cm,
 level 1/.style={sibling distance=2.7cm},
 every text node part/.style={font=\small},
 every lower node part/.style={font=\small}]
 \node[circle,draw] {$s_{init}$}
   child {
     node[circle,draw] {$s_{lb}'$} 
     {child[grow=down] {
       node[circle,draw] {$s_{lb}'$} 
       child[grow=down] {
         node[circle,draw] {$s_{lb}'$}
         edge from parent node[left] {$(i,i,i)$} 
                         }
       edge from parent node[left] {$(i,i,i)$}
                       }}
     edge from parent[dashed]
          }
   child {
     node[circle,draw] {$s_{a_1}$} 
     child[grow=down] {
       node[circle,draw] {$s_{a_1}$} 
       child[grow=down] {
         node[circle,draw] {$s_{a_1}$}
         edge from parent node[left] {$(i,i,i)$} 
                         }
       edge from parent node[left] {$(i,i,i)$}
                       }
     edge from parent[dashed]
          }
   child {
     node[circle,draw] {$s_{tr}'$} 
     child[grow=down] {
       node[circle,draw] {$s_{tr}'$} 
       child[grow=down] {
         node[circle,draw] {$s_{tr}'$}
         edge from parent node[left] {$(i,i,i)$} 
                         }
       edge from parent node[left] {$(i,i,i)$}
                       }
     edge from parent[dashed]
          }
   child {
     node[circle,draw] {$s_{q,a_2}$}
     child[grow=down] {
       node[circle,draw] {$s_{a_2'}$} 
       child[grow=down] {
         node[circle,draw] {$s_{a_2'}$}
         edge from parent node[left] {$(i,i,i)$} 
                         }
       edge from parent node[left] {\small $((q,q',R),i,i)$}
                       }
     edge from parent[dashed]
          }     
   child {
     node[circle,draw] {$s_{tr}'$}
     child[grow=down] {
       node[circle,draw] {$s_{q,q',R}$} 
       child[grow=down] {
         node[circle,draw] {$s_{tr}'$}
         edge from parent node[left] {\small $(i,(q,q',R),i)$} 
                         }
       edge from parent node[left] {\small $((q,q',R),i,i)$}
                       }
     edge from parent[dashed]
          }      
   child {
     node[circle,draw] {$s_{a_3}$} 
     child[grow=down] {
       node[circle,draw] {$s_{a_3}$} 
       child[grow=down] {
         node[circle,draw] {$s_{q',a_3}$}
         edge from parent node[left] {\small $(i,(q,q',R),i)$} 
                         }
       edge from parent node[left] {\small $(i,i,i)$}
                       }
     edge from parent[dashed]
          };
\end{tikzpicture}
\caption{Simulation of $a_1qa_2a_3\ \Ra_M\ a_1a_2'q'a_3$}
\label{F3}                                
\end{figure}

A computation step $a_1qa_2a_3\ \Ra_M\ a_1a_2'q'a_3$
in the Turing machine is simulated by extending the computation 
tree in Figure \ref{F1} as in Figure \ref{F3}.
The synchronization between the fourth and fifth paths 
is possible because, as we will see, these paths are 
$\sim_1$-equivalent. 
Similarly, the
synchronization between the fifth and sixth paths is
possible because these paths are $\sim_2$-equivalent. 

The simulation represented in these two figures proceeds as 
follows: in the observable history corresponding to the path 
ending in $s_{q,a_2}$, the only possibility for agent 1 to 
put the system in a state which satisfies $ok$ at the next 
level is to take action $(q,q',R)$, which corresponds to the 
transition $\delta(q,a_2)=(q',a_2',R)$ in the Turing machine.
Due to identic observability for agent 1, the same action has 
to be played by agent 1 in the history which ends in state 
$s_{tr}'$ which is next to the right of state $s_{q,a_2}$.
The effect of this action in state $s_{tr}'$ (combined with 
an idle action for agent 2) is to bring the system in state 
$s_{q,q',R}$. In this state, it's upto agent 2 to try to 
satisfy $ok$ at the next step, and he can only do this by 
applying the action $(q,q',R)$. The effect of this action in 
state $s_{q,q',R}$ is to bring the system back in state $s_{tr}'$.
But the same action has to be played by agent 2 in the history 
which ends in state $s_{a_3}$ on level 3 of the tree, due to 
identical observability. This play will lead the system to state 
$s_{q',a_3}$.

On the other hand, in state $s_{a_1}$, in order to ensure $ok$,
both agents must play idle, which leaves the system in state $s_{a_1}$.
Identical observability will then ensure that agent 1 has to play idle 
also in state $s_{tr}'$ which is next to the right of state $s_{a_1}$, 
and agent 2 has to play idle in state $s_{lb}'$ on 3rd and 4th levels.

The effect of all these is that level 4 on this tree encodes the 
configuration $a_1q'a_2'a_3$, which results from applying the transition 
$\delta(q,a_2)=(q',a_2',R)$ to the configuration $a_1qa_2a_3$.
States $s_{gen}$ and $s_{tr}$ are used for ``creating'' all the 
nodes that simulate tape cells. In a computation tree which satisfies 
the goal $\Box ok$, these are the only states to have two sons.

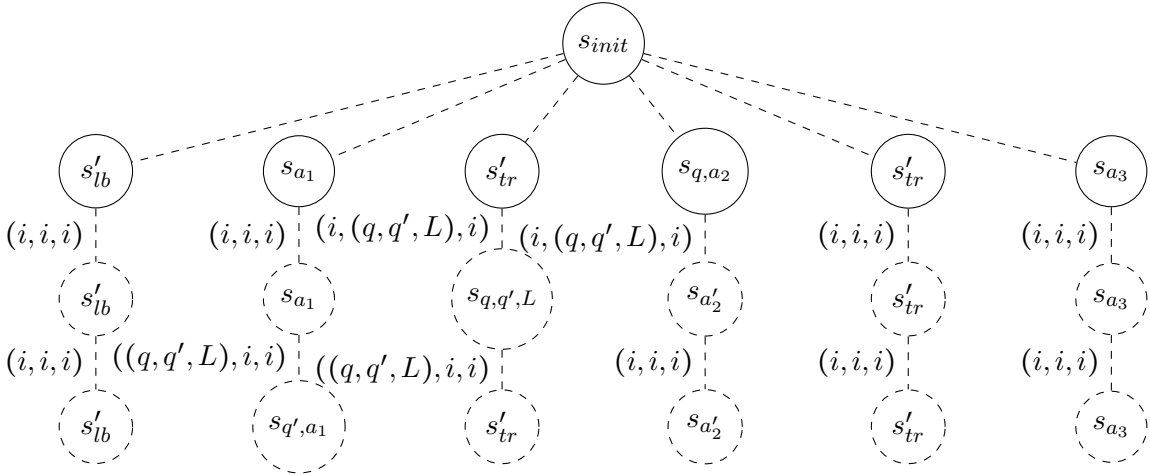
\begin{figure}[hbt]
\centering
\begin{tikzpicture}
[level distance=1.7cm,
 level 1/.style={sibling distance=2.7cm},
 every text node part/.style={font=\small},
 every lower node part/.style={font=\small}]
 \node[circle,draw] {$s_{init}$}
   child {
     node[circle,draw] {$s_{lb}'$} 
     child[grow=down] {
       node[circle,draw] {$s_{lb}'$} 
       child[grow=down] {
         node[circle,draw] {$s_{lb}'$}
         edge from parent node[left] {$(i,i,i)$} 
                         }
       edge from parent node[left] {$(i,i,i)$}
                       }
     edge from parent[dashed]
          }
   child {
     node[circle,draw] {$s_{a_1}$} 
     child[grow=down] {
       node[circle,draw] {$s_{a_1}$} 
       child[grow=down] {
         node[circle,draw] {$s_{q',a_1}$}
         edge from parent node[left] {$((q,q',L),i,i)$} 
                         }
       edge from parent node[left] {$(i,i,i)$}
                       }
     edge from parent[dashed]
          }
   child {
     node[circle,draw] {$s_{tr}'$} 
     child[grow=down] {
       node[circle,draw] {$s_{q,q',L}$} 
       child[grow=down] {
         node[circle,draw] {$s_{tr}'$}
         edge from parent node[left] {$((q,q',L),i,i)$} 
                         }
       edge from parent node[left] {\small $(i,(q,q',L),i)$}
                       }
     edge from parent[dashed]
          }
   child {
     node[circle,draw] {$s_{q,a_2}$}
     child[grow=down] {
       node[circle,draw] {$s_{a_2'}$} 
       child[grow=down] {
         node[circle,draw] {$s_{a_2'}$}
         edge from parent node[left] {$(i,i,i)$} 
                         }
       edge from parent node[left] {\small $(i,(q,q',L),i)$}
                       }
     edge from parent[dashed]
          }     
   child {
     node[circle,draw] {$s_{tr}'$}
     child[grow=down] {
       node[circle,draw] {$s_{tr}'$} 
       child[grow=down] {
         node[circle,draw] {$s_{tr}'$}
         edge from parent node[left] {\small $(i,i,i)$} 
                         }
       edge from parent node[left] {\small $(i,i,i)$}
                       }
     edge from parent[dashed]
          }      
   child {
     node[circle,draw] {$s_{a_3}$} 
     child[grow=down] {
       node[circle,draw] {$s_{a_3}$} 
       child[grow=down] {
         node[circle,draw] {$s_{a_3}$}
         edge from parent node[left] {\small $(i,i,i)$} 
                         }
       edge from parent node[left] {\small $(i,i,i)$}
                       }
     edge from parent[dashed]
          };
\end{tikzpicture}
\caption{Simulation of $a_1qa_2a_3\ \Ra_M\ q'a_1a_2'a_3$}
\label{F4}                                
\end{figure}

Figure \ref{F4} presents the simulation of the computation step
  $a_1qa_2a_3\ \Ra_M\ q'a_1a_2'a_3$
Note here that the r\^ole of agents 1 and 2 are interchanged because 
it is a left transition.

\begin{figure}[hbt]
\centering
\begin{tikzpicture}
[level distance=1.7cm,
 level 1/.style={sibling distance=5.4cm},
 level 2/.style={sibling distance=5cm},
 level 3/.style={sibling distance=4cm},
 level 4/.style={sibling distance=2cm},
 every text node part/.style={font=\small},
 every lower node part/.style={font=\small}]
 \node[circle,draw] {$s_{init}$}
   child {
     node[circle,draw] {$s_{init}'$}
     child {
       node[circle,draw] {$s_{lb}$}
       child {
         node[circle,draw] {$s_{lb}'$}
         child {
           node[circle,draw] {$s_{lb}'$}
           child {
             node[circle,draw] {$s_{lb}'$}
             child {
               node[circle,draw] {$s_{lb}'$}
               child {
                 node[circle,draw] {$s_{lb}'$}
                 edge from parent node[midway,left] {$(i,i,i)$}
                      }
               edge from parent node[midway,left] {$(i,i,i)$}
                    }
             edge from parent node[midway,left] {$(i,i,i)$}
                  }
           edge from parent node[midway,left] {$(i,i,i)$}
                }
         edge from parent node[midway,left] {$(i,(q_0),i)$}
              }
       edge from parent node[midway,left] {$(i,i,i)$}
            } 
     edge from parent node[midway,left] {$(i,i,br_1)$}
          }
   child {
     node[ellipse,draw] {$s_{gen}\,|\,p_1$} 
     child {
       node[circle,draw] {$s_B$}
       child {
         node[circle,draw] {$s_{q_0,B}$}
         child {
           node[circle,draw] {$s_a$}
           child {
             node[circle,draw] {$s_a$}
             child {
               node[circle,draw] {$s_a$}
               child {
                 node[circle,draw] {$s_{q_2,a}$}
                 edge from parent node[midway,left] {$((q_1,q_2,L),i,i)$}
                      }
               edge from parent node[midway,left] {$(i,i,i)$}
                    }
             edge from parent node[midway,left] {$(i,i,i)$}
                  }
           edge from parent node[midway,left] {$((q_0,q_1,R),i,i)$}
                }
         edge from parent node[midway,left] {$(i,(q_0),i)$}
              }
       edge from parent node[midway,left] {$(i,i,br_1)$}
            } 
     child {
       node[ellipse,draw] {$s_{tr}\,|\,p_2$}
       child {
         node[circle,draw] {$s_{tr}'$}
         child {
           node[circle,draw] {$s_{q_0,q_1,R}$}
           child {
             node[circle,draw] {$s_{tr}'$}
             child {
               node[circle,draw] {$s_{q_1,q_2,L}$}
               child {
                 node[circle,draw] {$s_{tr}'$}
                 edge from parent node[midway,left] {$((q_1,q_2,L),i,i)$}
                      }
               edge from parent node[midway,left] {$(i,(q_1,q_2,L),i)$}
                    }
             edge from parent node[midway,left] {$(i,(q_0,q_1,R),i)$}
                  }
           edge from parent node[midway,left] {$((q_0,q_1,R),i,i)$}
                }
         edge from parent node[midway,left] {$(i,i,br_1)$}
              }
       child { 
         node[ellipse,draw] {$s_{gen}\,|\,p_1$}
         child {
           node[circle,draw] {$s_B$}
           child {
             node[circle,draw] {$s_{q_1,B}$}
             child {
               node[circle,draw] {$s_b$}
               child {
                 node[circle,draw] {$s_b$}
                 edge from parent node[midway,left] {$(i,i,i)$}
                      }
               edge from parent node[midway,left] {$(i,(q_1,q_2,L),i)$}
                    }
             edge from parent node[midway,left] {$(i,(q_0,q_1,R),i)$}
                  }
           edge from parent node[midway,left] {$(i,i,br_1)$}
                }
         edge from parent node[midway,right] {$(i,i,br_2)$}
              }
       edge from parent node[midway,right] {$(i,i,br_2)$}
            }     
     edge from parent node[midway,right] {$(i,i,br_2)$}
          };
\end{tikzpicture}
\caption{Simulation of the computation $q_0B\ \Ra_M\ aq_1B\ \Ra_M\ q_2ab$. \label{fig5}}
\label{F5}                                
\end{figure}
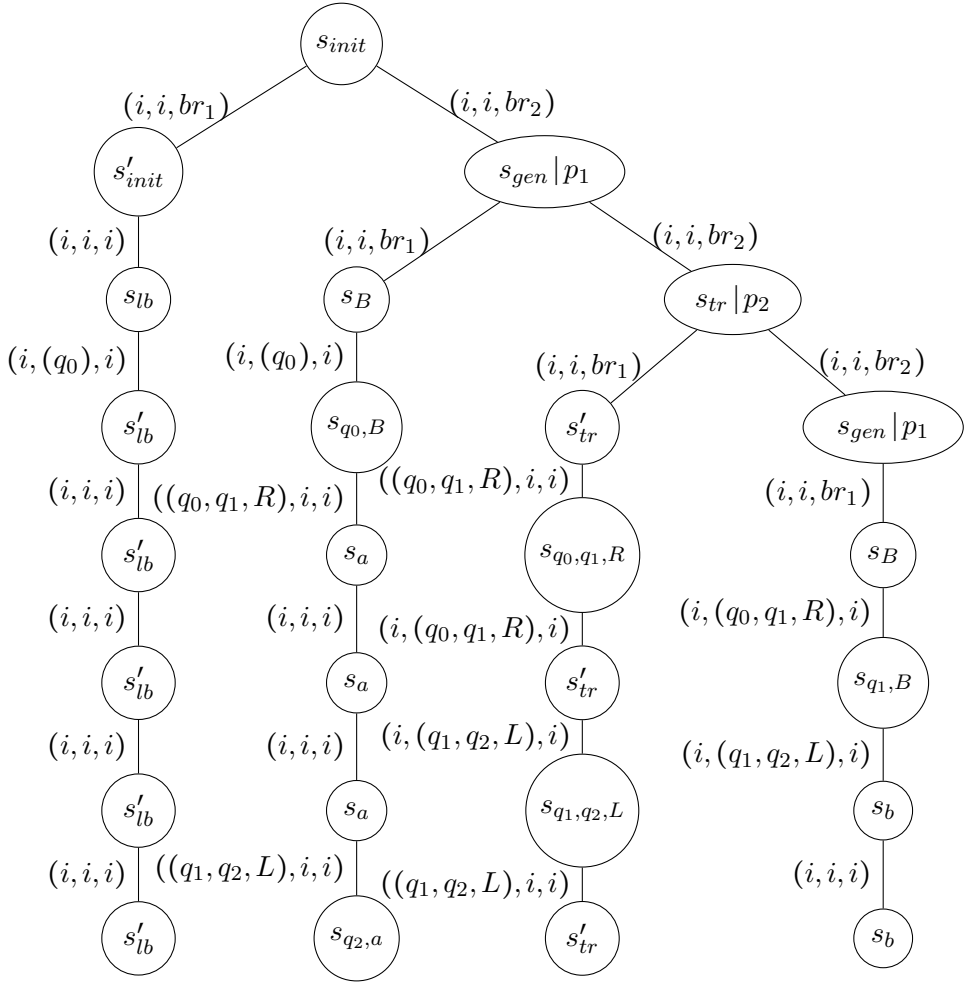

And in Figure \ref{F5}, a simulation of the computation 
  $q_0B\ \Ra_M\ aq_1B\ \Ra_M\ q_2ab$
is shown.

\paragraph{Construction of a game structure associated to $M$}

The concurrent game structure under imperfect information
$\cG=(Ag,S,\Pi,\pi,Act,(\sim_i|i\in Ag),d,\ra)$ that simulates the
deterministic Turing machine $M$ is based on three agents, i.e. 
$Ag=\{1,2,3\}$. 
Its set $S$ of states, together with their meaning, 
consists of:
\begin{itemize}
\item $s_{init}$  (the initial state);
\item $s_{init}'$ (copy of $s_{init}$);
\item $s_{lb}$ (specifies the left border of $M$'s tape);
\item $s_{lb}'$ (copy of $s_{lb}$);
\item $s_{gen}$ (initiates the generation of a new
  blank cell of $M$'s tape);
\item $s_{tr}$ (initiates the generation of a new cell separator);
\item $s_{tr}'$ (used for transferring information between to equivalent runs);
\item $s_a$, for any $a\in\Sigma$ (specifies that some tape cell holds $a$);
\item $s_{q,a}$, for any state $q\in Q$ and $a\in \Sigma$
  (specifies that $M$ is in state $q$ and the read/write head points 
   a cell holding symbol $a$);
\item $s_{q,q',X}$, for any $q,q'\in Q$ and $X\in\{L,R\}$ such
  that $\delta(q,a)=(q',a',X)$ for some $a$ and $a'$
  (specifies that the machine $M$ enters state $q'$ from 
  state $q$ by an $X$-move);
\item $s_{err}$ (``error'' state used to collect all ``unwanted'' 
  transitions the agents must avoid bringing the system in this state). 
\end{itemize}

The set of atomic propositions is $\Pi=\{p_1,p_2,ok\}$ and the labeling 
function $\pi$ is:
$$
\pi(s)=\left\{ 
    \begin{array}{ll}
      \{ok\},     & \mbox{if $s\in S-\{s_{gen},s_{tr},s_{err}\}$} \\
      \{p_1,ok\}, & \mbox{if $s=s_{gen}$} \\
      \{p_2,ok\}, & \mbox{if $s=s_{tr}$} \\
      \emptyset,  & \mbox{if $s=s_{err}$} 
    \end{array}
       \right.
$$
For the sake of simplicity, all states but $s_{err}$ will
be called {\em $ok$-states} (being labeled by $ok$). 

The relation $\sim_3$ is the identity. 
The equivalence relations $\sim_1$ and $\sim_2$ are defined by
  $$s\,\sim_i\,s'\ \ \ \text{iff}\ \ \ 
    (p_i\in \pi(s)\ \ \Lra\ \ p_i\in \pi(s')),$$
for any $i=1,2$. That is, $s$ and $s'$ are $\sim_i$-equivalent
if the agent $i$ observes $p_i$ either in both states $s$ 
and $s'$ or in none of them. 

The set $Act$ of actions consists of:
\begin{itemize}
\item {\em idle}, which is meant to say that the agent doing 
  it is not ``in charge of'' accomplishing some local objective
  (this action will be abbreviated by $i$ in our pictures
  and whenever no confusion may arise);
\item $(q_0)$, which is an action meant to set up the initial
  state of $M$;
\item $(q,q',X)$, for any $q,q'\in Q$ and $X\in\{L,R\}$ with
  $\delta(q,a)=(q',a',X)$ for some $a,a'\in\Sigma$. 
  Such an action simulates the passing of $M$ from $q$ to $q'$
  by an $X$-move;
\item $br_1$ and $br_2$, which are two ``branching'' actions. 
\end{itemize}
The agents $1$ and $2$ are allowed to perform any action but 
$br_1$ and $br_2$, while the third agent can only perform $br_1$,
$br_2$, and $idle$. More precisely, $d(i,s)=Act-\{br_1,br_2\}$ 
for any $i\in\{1,2\}$ and state $s$, $d(3,s)=\{br_1,br_2\}$ if
$s\in\{s_{init},s_{gen},s_{tr}\}$, and $d(3,s)=idle$, otherwise.

Note that the agents' actions are designed such that 
$d(i,s)=d(i,s')$ for any agent $i$ and 
states $s$ and $s'$ with $s\,\sim_i\,s'$.

The transition relation of the game structure is as follows:
\begin{itemize}
\item $s_{init} \sdup{(i,i,br_1)} s_{init}'$ and   
      $s_{init} \sdup{(i,i,br_2)} s_{gen}$ and
      $s_{init} \sdup{c} s_{err}$, for any $c$
      different from the above action tuples;
\item $s_{init}' \sdup{(i,i,i)} s_{lb}$ and 
      $s_{init}' \sdup{c} s_{err}$, for any $c\not=(i,i,i)$;
\item $s_{lb} \sdup{(i,(q_0),i)} s_{lb}'$ and 
      $s_{lb} \sdup{c} s_{err}$, for any $c\not=(i,(q_0),i)$;
\item $s_{lb}' \sdup{(i,i,i)} s_{lb}'$ and 
      $s_{lb}' \sdup{c} s_{err}$, for any $c\not=(i,i,i)$;
\item $s_{gen} \sdup{(i,i,br_1)} s_B$ and 
      $s_{gen} \sdup{(i,i,br_2)} s_{tr}$ and  
      $s_{gen} \sdup{c} s_{err}$, for any $c$
      different from the above action tuples;
\item $s_{tr} \sdup{(i,i,br_1)} s_{tr}'$ and 
      $s_{tr} \sdup{(i,i,br_2)} s_{gen}$ and  
      $s_{tr} \sdup{c} s_{err}$, for any $c$
      different from the above action tuples;
\item for any $a\in\Sigma$, the transitions at $s_a$ are:
  \begin{itemize}
  \item $s_a \sdup{(i,i,i)} s_a$;
  \item $s_B \sdup{(i,(q_0),i)} s_{q_0,B}$;
  \item $s_a \sdup{(i,(q,q',R),i)} s_{q,a}$, for any action
    $(q,q',R)$;
  \item $s_a \sdup{((q,q',L),i,i)} s_{q,a}$, for any action
    $(q,q',L)$;
  \item $s_a \sdup{c} s_{err}$, for any $c$ different from 
    any of the above actions;
  \end{itemize}
\item for any $q\in Q$ and $a\in\Sigma$, the transitions at 
  $s_{q,a}$ are:
  \begin{itemize}
  \item $s_{q,a} \sdup{((q,q',R),i,i)} s_{a'}$, if 
              $\delta(q,a)=(q',a',R)$;
  \item $s_{q,a} \sdup{((q,q',L),i,i)} s_{a'}$, if 
              $\delta(q,a)=(q',a',L)$;
  \item $s_{q,a} \sdup{c} s_{err}$, for any $c$ different from 
    any of the above actions;
  \end{itemize}
\item the transitions at $s_{tr}'$ are:
  \begin{itemize}
  \item $s_{tr}' \sdup{(i,i,i)} s'_{tr}$.
  \item $s_{tr}' \sdup{((q,q',R),i,i)} s_{q,q',R}$, for any action
    $(q,q',R)$;
  \item $s_{tr}' \sdup{(i,(q,q',L),i)} s_{q,q',L}$, for any action
    $(q,q',L)$;
  \item $s_{tr}' \sdup{c} s_{err}$, for any $c$ different from 
    any of the above actions;
  \end{itemize}
\item $s_{q,q',R} \sdup{(i,(q,q',R),i)} s_{tr}'$ and 
      $s_{q,q',L} \sdup{((q,q',L),i,i)} s_{tr}'$ and 
      $s_{q,q',X} \sdup{c} s_{err}$, for any $X$ and
      any $c$ different from any of the above actions.
\end{itemize}

\paragraph{Proof of the correctness of the construction}

Let $M$ be a deterministic Turing machine. 
Without loss of generality we may assume that $M$, starting 
in state $q_0$, will never reach again $q_0$. 

First, we prove that if $M$ does not halt on the empty word
then $(\cG,s_{init})\models_{iR}\ceggg[\{1,2\}]ok$.
According to Remark \ref{R1}, it suffices to show that,
if $M$ does not halt on the empty word, then there exists
a strategy $\sigma=(\sigma_1,\sigma_2)$ for the agents $1$
and $2$ in $\cG$ such that any $s_{init}$-rooted computation 
tree of $\cG$ under $\sigma$ has only nodes labeled by 
$ok$-states.

In order to define $\sigma$ with the property above, 
we classify the non-empty sequences of states of $\cG$ as 
follows:
\begin{itemize}
\item a sequence $\alpha\in S^+$ is of {\em type 1} if 
   $\alpha=s_{init}s_{init}'\alpha'$, where $\alpha'\in S^*$;
\item a sequence $\alpha\in S^+$ is of {\em type 2} if 
   $\alpha=s_{init}s_{gen}\alpha'$, where $\alpha'\in S^*$.
   Type 2 sequences of states can be further classified 
   according to the number of states $s_{gen}$ and $s_{tr}$ 
   they contain:
   \begin{itemize} 
   \item a sequence $\alpha$ is of {\em type $2(i)(i-1)$}, 
     where $i\geq 1$, if 
     $\alpha=s_{init}(s_{gen}s_{tr})^{i-1}s_{gen}\alpha'$,
     where $\alpha'\in S^*$ does not contain $s_{gen}$
     and $s_{tr}$;
   \item a sequence $\alpha$ is of {\em type $2(i)(i)$}, 
     where $i\geq 1$, if 
     $\alpha=s_{init}(s_{gen}s_{tr})^{i}\alpha'$,
     where $\alpha'\in S^*$ does not contain $s_{gen}$
     and $s_{tr}$.
   \end{itemize}
\end{itemize}
Of course, there are sequences $\alpha\in S^+$ which are 
neither of type 1 nor of type 2. 
A path $\tau$ of a computation tree of $\cG$ will be called
of {\em type $x$} if $l_1(\tau)$ is of type $x$, where $x$ is
as above.

The following claim follows easily from definitions.

\begin{claim}\label{C1}
Let $\alpha$ and $\alpha'$ be two non-empty sequences of
states. Then, the following properties hold:
\begin{enumerate}
\item If $\alpha$ is of type 1 and $\alpha'$ is of type 2, 
  then $\alpha \not \sim_1 \alpha'$;
\item  If $\alpha$ is of type 1 and $\alpha'$ is of type 2
  and $\alpha\sim_2 \alpha'$, then $\alpha'$ is
  of type $2(1)(0)$;
\item If $\alpha$ and $\alpha'$ are of type 2, have a different
  number of $s_{gen}$ or $s_{tr}$ states, and 
  $\alpha \sim_1 \alpha'$, then $\alpha$ is of type $2(i)(i-1)$ 
  and $\alpha'$ is of type $2(i)(i)$, or vice-versa;
\item If $\alpha$ and $\alpha'$ are of type 2, have a different
  number of $s_{gen}$ or $s_{tr}$ states, and 
  $\alpha \sim_2 \alpha'$, then $\alpha$ is of type $2(i)(i)$ 
  and $\alpha'$ is of type $2(i+1)(i)$, or vice-versa.
\end{enumerate}
\end{claim}

Now, define a strategy $\sigma=(\sigma_1,\sigma_2)$ as follows:
\begin{itemize}
\item $\sigma_1(s_{init})=\sigma_1(\alpha)=idle$,
  for any type 1 sequence $\alpha\in S^+$;
\item $\sigma_2(s_{init})=\sigma_2(\alpha)=idle$, 
  for any type 1 sequence $\alpha\in S^+$ different from 
  $s_{init}s_{init}'s_{lb}$, and 
  $\sigma_2(s_{init}s_{init}'s_{lb})=(q_0)$;
\item $\sigma_1(\alpha s_{q,a})=(q,q',R)=\sigma_1(\alpha' s_{tr}')$, 
  for any $\alpha s_{q,a}$ of type $2(i)(i-1)$ and any
  $\alpha' s_{tr}'$ of type $2(i)(i)$ for which $i\geq 1$ 
  and the following property holds:
  \begin{itemize}
  \item $|\alpha s_{q,a}|=3+(2j-1)=|\alpha' s_{tr}'|$ for some 
    $j\geq 1$, and
    the agent $1$ simulating the first $j$ steps of $M$
    deduces that the current configuration of $M$ is of
    the form $uqav$, where $|u|=i-1$, and 
    $\delta(q,a)=(q',a',R)$, for some $q'$ and $a'$; 
  \end{itemize}
\item $\sigma_1(\alpha s_{a})=(q,q',L)=\sigma_1(\alpha' s_{q,q',L})$, 
  for any $\alpha s_{a}$ of type $2(i)(i-1)$ and any
  $\alpha' s_{q,q',L}$ of type $2(i)(i)$ for which $i\geq 1$ 
  and the following property holds:
  \begin{itemize}
  \item $|\alpha s_{a}|=3+2j=|\alpha' s_{q,q',L}|$ for some 
    $j\geq 1$, and
    the agent $1$ simulating the first $j$ steps of $M$
    deduces that the current configuration of $M$ is of
    the form $uaqbv$, where $|u|=i-1$, and 
    $\delta(q,b)=(q',b',L)$, for some $q'$ and $b'$;
  \end{itemize}
\item $\sigma_2(\alpha s_{q,q',R})=(q,q',R)=\sigma_2(\alpha' s_{a})$, 
  for any $\alpha s_{q,q',R}$ of type $2(i)(i)$ and
  any $\alpha' s_{a}$ of type $2(i+1)(i)$ for which $i\geq 1$ 
  and the following property holds:
  \begin{itemize}
  \item $|\alpha s_{q,q',R}|=3+2j=|\alpha' s_{a}|$ for some 
    $j\geq 1$, and
    the agent $2$ simulating the first $j$ steps of $M$
    deduces that the current configuration of $M$ is of
    the form $uqav$, where $|u|=i-1$, and 
    $\delta(q,a)=(q',a',R)$, for some $q'$ and $a'$; 
  \end{itemize}
\item $\sigma_2(\alpha s_{tr}')=(q,q',L)=\sigma_2(\alpha' s_{q,a})$, 
  for any $\alpha s_{tr}'$ of type $2(i)(i)$ and any
  $\alpha' s_{q,a}$ of type $2(i+1)(i)$ for which $i\geq 1$ 
  and the following property holds:
  \begin{itemize}
  \item $|\alpha s_{tr}'|=3+(2j-1)=|\alpha' s_{q,a}|$ for some 
    $j\geq 1$, and
    the agent $2$ simulating the first $j$ steps of $M$
    deduces that the current configuration of $M$ is of
    the form $uaqbv$, where $|u|=i-1$, and 
    $\delta(q,b)=(q',b',L)$, for some $q'$ and $b'$;
  \end{itemize}
\item $\sigma_2(s_{init}s_{gen}s_B)=(q_0)$;
\item $\sigma_1(\alpha)=idle$ and $\sigma_2(\alpha')=idle$
  for all the other cases.
\end{itemize}

The strategies $\sigma_1$ and $\sigma_2$ are both compatible
with $d$, $\sigma_1$ is compatible with $\sim_1$, and 
$\sigma_2$ is compatible with $\sim_2$. 

Any tree with exactly one node (its root) labeled
by $s_{init}$ is an $s_{init}$-rooted computation tree of
$\cG$ under $\sigma$ and its nodes are all labeled by 
$ok$-states. 

Assume that $\cT$ is an $s_{init}$-rooted computation tree
of $\cG$ under $\sigma$ and all its nodes are labeled by
$ok$-states. 
It is easy to see that $\cT$ may only have type 1, type
$2(i)(i-1)$, or type $2(i)(i)$ paths, for some $i\geq 1$.
Any extension $\cT'$ of $\cT$ (i.e., $\cT\Ra_{\cG,\sigma}\cT'$)
adds new nodes to $\cT$ which cannot be labeled by $s_{err}$
because $M$ does not halt (see the definition of $\sigma$). 
Therefore, any $s_{init}$-rooted computation tree of $\cG$
under $\sigma$ has all its nodes labeled by $ok$-states.

\bigskip
Conversely, we show that $M$ does not halt on the empty 
word if all $s_{init}$-rooted computation trees of $\cG$
under some strategy $\sigma$ for $\{1,2\}$ have only nodes
labeled by $ok$-states.  

Let $\sigma$ be a strategy with the property above and consider 
an $s_{init}$-rooted computation tree $\cT=(V,E,v_0,l_1,l_2)$ 
under $\sigma$.
A node $v$ of $\cT$ will be called of {\em type $x$} if
$l_1(path_\cT(v_0,v))$ is of type $x$ ($x$ is 1, 2, $2(i)(i-1)$,
or $2(i)(i)$, for some $i\geq 1$). 

We then define a partial ordering $\prec_\cT$ on the nodes
of $\cT$ as the least partial ordering with the following 
properties:
\begin{itemize}
 \item if $v$ and $v'$ are nodes on the same level of $\cT$ and 
       $l_1(v')\in\{s_{gen},s_{tr}\}$, then $v\prec_\cT v'$;
 \item if $v$ and $v'$ are nodes on the same level of $\cT$ and 
       there exist $u$ on the path from root to $v$ and 
       $u'$ on the path from root to $v'$ with $u\prec_\cT u'$,
       then $v\prec_\cT v'$
\end{itemize}

Some properties of $\cT$ and its level sets are listed in 
the sequel.

\begin{claim}\label{C2}
Let $\cT=(V,E,v_0,l_1,l_2)$ be an $s_{init}$-rooted computation 
tree of $\cG$ under $\sigma$, and $n\geq 1$. Then:
\begin{enumerate}
\item $level_\cT(n)$ has at most $n+1$ nodes, and each of them
  is either of type 1, or of type 2, or of type
  $2(i)(i-1)$, or of type $2(i)(i)$, for some $i\geq 1$;
\item $level_\cT(n)$ contains at most one node of type 1;
\item $level_\cT(n)$ contains at most one node of type 
  $2(i)(i-1)$ and at most one node of type $2(i)(i)$, for
  each $i \leq \lceil n/2 \rceil $;
\item for any $v,v'\in level_\cT(n)$, $v\prec_\cT v'$ if
  and only if one of the following properties hold:
  \begin{enumerate}
  \item $v=v'$;
  \item $v$ is of type $1$;
  \item $v$ is of type $2(i)(i')$, $v'$ is of type $2(j)(j')$,
    and $i<j$ or, if $i=j$ then $i'<j'$.
  \end{enumerate}
\item $\prec_\cT$ is a total ordering on $level_\cT(n)$.
\end{enumerate}
\end{claim}
\begin{pf}
All the properties in Claim \ref{C2} can be proved by induction 
on $n\geq 1$ and make use of the fact that all nodes of $\cT$ 
are labeled by $ok$-states. Thus, if $v$ is a node on the 
level $n$ of $\cT$ and it is not label by $s_{gen}$ or $s_{tr}$, 
then it may have at most one descendant $v'$ on the level $n+1$
(by $\sigma$, each of the agents $1$ and $2$ has exactly one 
choice at $l_1(v)$, and by $d_3$, the agent $3$ has exactly
one choice as well at $l_1(v)$). 
Moreover, $v'$ and $v$ have the same type. If $v$ is labeled
by $s_{gen}$, then its type is $2(i)(i-1)$ for some $i\geq 1$,
and it may have at most two descendants $v'$ and $v''$ on the 
level $n+1$ (by $\sigma$, each of the agents $1$ and $2$ has 
exactly one choice at $l_1(v)$, but the agent $3$ has two
choices).  
One of this descendants is of type $2(i)(i-1)$, while
the other is of type $2(i)(i)$ and it is labeled by $s_{tr}$. 
Similarly, if $v$ is labeled by $s_{tr}$, then its type is 
$2(i)(i)$ for some $i\geq 1$, and it may have at most two 
descendants $v'$ and $v''$ on the level $n+1$. One of this 
descendants is of type $2(i)(i)$, while the other is of type 
$2(i+1)(i)$ and it is labeled by $s_{gen}$. 

Combining these remarks with the fact that $level_\cT(1)$ may
contain at most two nodes, one of them labeled by $s_{init}'$
(which is of type 1) and the other by $s_{gen}$, we obtain 
(1), (2), and (3) in the Claim. 

(4) follows from the definition of $\prec_\cT$ and the
above properties, and (5) follows from (4).
\qed
\end{pf}

If $level_\cT(n)=\{v_1,\ldots,v_{n+1}\}$ of an 
$s_{init}$-rooted computation tree $\cT$ of $\cG$ under $\sigma$ 
has exactly $n+1$ nodes, then we say that it is {\em complete}. 
Moreover, if we assume that $v_1\prec_\cT\cdots\prec_\cT v_{n+1}$, 
then we may view $level_\cT(n)$ as a sequence of nodes,
$v_1\cdots v_{n+1}$.

\begin{claim}\label{C3}
Let $\cT=(V,E,v_0,l_1,l_2)$ be an $s_{init}$-rooted computation 
tree of $\cG$ under $\sigma$, and $n\geq 1$ such that 
$level_\cT(n)$ is complete and its sequence of nodes is
$v_1\cdots v_{n+1}$. Then, the following properties hold:
\begin{enumerate}
\item $level_\cT(m)$ is complete, for any $m\leq n$;
\item $v_1$ is of type $1$, $v_{2i}$ is of type $2(i)(i-1)$,
  and $v_{2i+1}$ is of type $2(i)(i)$, for all $i\geq 1$ 
  with $2i\leq n$;
\item 
  \begin{enumerate}
  \item $l_1(path_\cT(v_0,v_1))\sim_2 l_1(path_\cT(v_0,v_2))$;
  \item $l_1(path_\cT(v_0,v_{2i}))\sim_1 l_1(path_\cT(v_0,v_{2i+1}))$, 
        for all $i\geq 1$ with $2i\leq n$;
  \item $l_1(path_\cT(v_0,v_{2i+1}))\sim_2 l_1(path_\cT(v_0,v_{2(i+1)}))$,
        for all $i\geq 1$ with $2i+1\leq n$;
  \end{enumerate}
\item $l_1(v_1\cdots v_{n+1})$ is of the one of the following forms:
  \begin{enumerate}
  \item $s_{init}'s_{gen}$, if $n=1$;
  \item $s_{lb}s_Bs_{tr}$, if $n=2$;
  \item $s_{lb}'s_{a_1}s_{tr}'\cdots 
         s_{a_{j-1}}s_{tr}'s_{q,a_j}s_{tr'}s_{a_{j+1}}\cdots 
         s_{tr}'s_{a_m}s_{tr}'s_{gen}$, if $n>2$ is odd, 
      where $a_1,\ldots,a_m\in\Sigma$, $q\in Q$, $m=(n-1)/2$,
      and $1\leq j\leq m$ (for $j=1$, $s_{a_1}$ becomes
      $s_{q,a_1}$, and for $j=m$, $a_m$ becomes $s_{q,a_m}$);
  \item $s_{lb}'s_{a_1}s_{tr}'\cdots 
         s_{a_{j-1}}s_{tr}'s_{a_j}s_{q,q',X}s_{a_{j+1}}\cdots 
         s_{tr}'s_{a_{m-1}}s_{tr}'s_Bs_{tr}$, if $n>2$ is even, 
      where $a_1,\ldots,a_{m-1}\in\Sigma$, $q,q'\in Q$, $X\in\{L,R\}$,
      $m=n/2$, and $1\leq j\leq m-1$;
  \end{enumerate}
\item there exists an $s_{init}$-rooted computation tree 
  $\cT'$ of $\cG$ under $\sigma$ such that 
  $\cT\stackrel{*}{\Ra}_{\cG,\sigma}\cT'$
  and $level_{\cT'}(n+1)$ is complete. 
  Moreover, if the sequence of nodes of $level_\cT(n)$ has the form
  (4a) ((4b), (4c), (4d)), then $level_{\cT'}(n+1)$ has the form
  (4b) ((4c), (4d), (4c), respectively).
\end{enumerate}
\end{claim}
\begin{pf}
(1), (2), and (3) can be proved in a similar way to the statements 
in Claim \ref{C2}. 

We prove (4) and (5) together. 
It is easy to show that $l_1(v_1\cdots v_{n+1})$ has the form
(4a) if $n=1$.
As $l_1(path_\cT(v_0,v_1))\sim_2 l_1(path_\cT(v_0,v_2))$
and $\cT$ has only $ok$-states, the strategy $\sigma_2$
should select only $idle$ as the only choice for agent 2 
at $l_1(v_1)$ and $l_1(v_2)$. 
$\sigma_1$ should select $idle$ for agent 1 at $l_1(v_1)$
and $l_1(v_2)$, while the agent 3 has the only choice $idle$
at $l_1(v_1)$ and two choices, $br_1$ and $br_2$, at $l_1(v_2)$.
Therefore, we can extend $\cT$ by adding a new descendant $v_1'$ 
of $v_1$ and two new descendants $v_2'$ and $v_2''$ of $v_2$,
by the rules
  $$l_1(v_1)\sdup{(i,i,i)} l_1(v_1')=s_{lb},\ \ \ 
    l_1(v_2)\sdup{(i,i,br_1)} l_1(v_2')=s_B,\ \ \
    l_1(v_2)\sdup{(i,i,br_2)} l_1(v_2'')=s_{tr}.$$
We obtain a new $s_{init}$-rooted computation tree 
$\cT'$ of $\cG$ under $\sigma$ whose level 2 satisfies
(4) and (5). 

Assume $n=2$ and $l_1(v_1,v_2,v_3)=s_{lb}s_Bs_{tr}$.
As $l_1(path_\cT(v_0,v_1))\sim_2 l_1(path_\cT(v_0,v_2))$
and $\cT$ has only $ok$-states, the strategy $\sigma_2$
should select only $(q_0)$ as the only choice for agent 2 
at $l_1(v_1)$ and $l_1(v_2)$. 
The agents 1 has the only choice $idle$ at $l_1(v_1)$ and 
$l_2(v_2)$ (by $\sigma_1$), and the agent 3 has the same 
choice at these states (by $d_3$). Therefore, we can add
a new descendant $v_1'$ of $v_1$ and a new descendant $v_2'$
of $v_2$ by the rules 
  $$l_1(v_1)=s_{lb}\sdup{(i,(q_0),i)} l_1(v_1')=s_{lb}'
    \mbox{\ \ \ and\ \ \ }
    l_1(v_2)=s_B\sdup{(i,(q_0),i)} l_1(v_2')=s_{q_0,B}.$$
There are two choices at $l_1(v_3)$, namely $(i,i,br_1)$ and
$(i,i,br_2)$, allowing to add two descendants $v_3'$ and
$v_3''$ of $v_3$ on the next level. 
Moreover, $l_1(v_3')=s_{tr}'$ and $l_1(v_3'')=s_{gen}$.
As a conclusion, $\cT$ can be extended to a new tree $\cT'$
whose sequence of nodes on level 3 are 
$v_1'v_2'v_3'v_3''$ and 
$l_1(v_1'v_2'v_3'v_3'')=s_{lb}'s_{q_0,B}s_{tr}'s_{gen}$
which is the form (4c). Moreover, (5) holds too. 

Assume $n>2$ odd, $l_1(v_1\cdots v_{n+1})$ of the form (4c),
and $j>1$ (the case $j=1$ can be discussed in a similar way). 
We have that $l_1(v_{2j})=s_{q,a_j}$ and 
$l_1(v_{2j-1})=l_1(v_{2j+1})=s_{tr}'$.
Due to the fact that 
$l_1(path_\cT(v_0,v_{2j}))\sim_1 l_1(path_\cT(v_0,v_{2j+1}))$
and $\cT$ has only $ok$-states, $\sigma_1$ should select an
action of the form $(q,q',R)$ or $(q,q',L)$ for agent 1
as a choice at $l_1(v_{2j})$ and $l_1(v_{2j+1})$
($q'\in Q$ and this choice is obtained from the transition 
function of $M$). Assume that this choice is $(q,q',R)$
and $\delta(q,a_j)=(q',a_j',R)$ 
(the other case is similar to this). Each of the agents
2 and 3 has exactly one choice at $l_1(v_{2j})$ and 
$l_1(v_{2j+1})$, namely $idle$. Therefore, $\cT$ can be
extended by adding two new descendants $v_{2j}'$ and
$v_{2j+1}'$ by the rules 
  $$l_1(v_{2j})=s_{q,a_j}\sdup{((q,q',R),i,i)} l_1(v_{2j}')=s_{a_j'}
    \mbox{\ \ \ and\ \ \ }
    l_1(v_{2j+1})=s_{tr}'\sdup{((q,q',R),i,i)} l_1(v_{2j+1}')=s_{q,q',R}.$$
For the nodes $v_i$ with $i\not\in\{2j,2j+1,n+1\}$, there is 
exactly one choice for each agent, namely $idle$, and therefore,
a new descendant $v_i'$ of $v_i$ can be added by the rule 
  $$l_1(v_i)\sdup{(i,i,i)}l_1(v_i')=l_1(v_i).$$
For the node $v_{n+1}$ we may reason as in the case $n=2$ above.
Two descendants $v_{n+1}'$ and $v_{n+1}''$ can be added,
with $l_1(v_{n+1}')=s_B$ and $l_1(v_{n+1}'')=s_{tr}$.

In this way, we obtain a new tree $\cT'$ whose level $n+1$
satisfies (4) and (5). 

The case ``$n>2$ even and $l_1(v_1\cdots v_{n+1})$ of the form (4d)''
can be treated analogously to the above one. 
\qed
\end{pf}

Consider further the homomorphism $h:S\ra (Q\cup\Sigma)^*$ given by:
$$
 h(s) = \begin{cases} 
          a,       & \mbox{if } s=s_a \\
          qa,      & \mbox{if } s=s_{q,a} \\
          \lambda, & \mbox{otherwise}
        \end{cases}
$$

We shall write $h(level_\cT(n))$ for $h(v_1\cdots v_{n+1})$,
where $v_1\cdots v_{n+1}$ is the sequence of nodes associated
to complete level $level_\cT(n)$ of some $s_{init}$-rooted 
computation tree $\cT$ of $\cG$ under $\sigma$.

\begin{claim}\label{C4}
Let $\cT=(V,E,v_0,l_1,l_2)$ be an $s_{init}$-rooted computation 
tree of $\cG$ under $\sigma$, and $n\geq 3$ odd such that 
$level_\cT(n)$ is complete. Then:
\begin{enumerate}
\item $h(level_\cT(n))\in \Sigma^* Q\Sigma\Sigma^*$;
\item there exists an $s_{init}$-rooted computation tree 
  $\cT'$ of $\cG$ under $\sigma$ such that 
  $\cT\stackrel{*}{\Ra}_{\cG,\sigma}\cT'$, 
  $level_{\cT'}(n+2)$ is complete, and 
  $h(level_\cT(n))\Ra_M h(level_{\cT'}(n+2))$.
\end{enumerate}
\end{claim}
\begin{pf}
From the definition of $h$, Claim \ref{C3}, and by 
inspecting the proof of Claim \ref{C3}. 
\qed
\end{pf}

It is straightforward to see that there exists an
$s_{init}$-rooted computation tree $\cT$ of $\cG$ under $\sigma$
whose $level_\cT(3)$ is complete. Moreover, by Claim \ref{C3},
we have $h(level_\cT(3))=q_0B$ (that is, the initial configuration
of $M$). Then, combining with Claim \ref{C4}, we obtain that
$M$ does not halt on the empty word if 
all $s_{init}$-rooted computation trees of $\cG$
under some strategy $\sigma$ for $\{1,2\}$ have only nodes
labeled by $ok$-states.

\bigskip
Our discussion above leads to:

\begin{thm}
The model checking problem for $ATL_{iR}$ is undecidable.
\end{thm}

\section{Conclusions}

The proof above shows that the strategies used by the agents
1 and 2 to simulate the deterministic Turing machine $M$ are
primitive recursive. Therefore, the crucial elements which 
allow to simulate $M$ are the equivalence relations $\sim_1$
and $\sim_2$. These equivalence relations are ``inter-related''
and are used to transfer information from one computation path 
can be transferred to another computation path. 

A deeper analysis of the nature of the observational equivalence
relations associated to agents in a $CGS$ would be interesting.


\end{document}